%% file: main.tex
\documentclass[twocolumn, amsmath,amssymb, aps, pre]{revtex4-1}

\usepackage{textcomp}
\usepackage[utf8]{inputenc}
\usepackage{amsmath}
\usepackage{mathtools}
\usepackage{graphicx}
\usepackage{caption}
\usepackage[colorlinks=false]{hyperref}
\usepackage{color}
\usepackage[euler]{textgreek}

\makeatletter
\def\convertto#1#2{\strip@pt\dimexpr #2*65536/\number\dimexpr 1#1}
\makeatother

\begin{document}
\title{Mechanosensitive Self-Assembly of Myosin II Minifilaments}
\author{Justin Grewe} \author{Ulrich S. Schwarz}\email{schwarz@thphys.uni-heidelberg.de}
\affiliation{Institute for Theoretical Physics and Bioquant, Heidelberg University, Heidelberg, Germany}

\date{\today}

\begin{abstract}
Self-assembly and force generation are two central processes in biological systems that usually are considered in separation. 
However, the signals that activate non-muscle myosin II molecular motors simultaneously lead to self-assembly into myosin II minifilaments as well as progression of the motor heads through the crossbridge cycle.
Here we investigate theoretically the possible effects of coupling these two processes. Our assembly model, which builds upon 
a consensus architecture of the minifilament, predicts a critical aggregation concentration at which the assembly kinetics slows 
down dramatically. The combined model predicts that increasing actin filament concentration and force both lead to a decrease in the critical aggregation concentration. We suggest that due to these effects, myosin II minifilaments in a filamentous context might be in a critical state that reacts faster to varying conditions than in solution. We finally compare our model
to experiments by simulating fluorescence recovery after photobleaching.
\end{abstract}

\maketitle
\input{intro}
\input{model}
\input{results}
\input{outlook}

\begin{acknowledgments}
The authors would like to thank Rasmus Schroeder, Tom Kaufmann and Falko Ziebert for helpful interactions.
This work was supported by the cluster of excellence Structures (EXC-2181/1-390900948, project CP3).
U.S.S. is also a member of the clusters of excellence CellNetworks and 3DMM2O, and of the
Interdisciplinary Center for Scientific Computing (IWR). 
\end{acknowledgments}
 

%

\input{supplement}

\end{document}

%% file: intro.tex
\section{Introduction}

Molecular motors powered by ATP-consumption are ubiquitous in living organisms, converting
chemical energy into movement and force at the right time and place \cite{HowardBook2001}. The most important
molecular motor for force generation is the two-headed non-processive molecular motor myosin II, which occurs in many different variants.
In skeletal muscle, hundreds of skeletal myosin II motors are assembled into the thick filament
that forms the core of the sarcomere. Large assemblies of the corresponding myosin II variants also
exist in cardiac and smooth muscle. In non-muscle cells, however, non-muscle myosin II
assembles into much smaller groups, so-called myosin II minifilaments, that
due to their small size can be dynamically regulated to generate forces on demand,
in particular in the actomyosin cortex and in stress fibers \cite{Vicente-Manzanares2009, Dasbiswas2018}.
Very importantly for the way myosin II minifilaments function, it is not only
force generation, but also assembly that is regulated in non-muscle cells. In particular,
the Rho-pathway leading to myosin II minifilament activation has two branches,
one regulating actin assembly through the formin mDia1 and one leading to
phosphorylation of the myosin II regulatory chain \cite{Ridley2015,Bershadsky2006}. This in turn leads both to
myosin II assembly and cycling of the motor heads.
Together, these different elements make sure that myosin II minifilaments are assembled
in a functional state in which motor heads and actin filaments work together synergistically.
However, because assembly and force generation of non-muscle myosin II minifilaments are usually
studied in isolation, no quantitative understanding exists for how these two processes are coupled in cells.
Here we introduce and analyze a mathematical model for this purpose.

Regarding force generation, we start from earlier models of force generation,
which occurs by myosin cycling through a set of mechanochemical states, as first formalized by Huxley \cite{Huxley1957}.
Briefly, myosin binds to actin, then the lever arm performs the powerstroke, the myosin detaches from actin and the lever arm resets.
This cycle is powered by ATP-hydrolysis and each of these states corresponds to a step in the hydrolysis cycle.
Cross-bridge models are master equation models using this discrete sets of states, and have
been used with great success to study a variety of effects that arise due to the mechanochemistry of molecular
motors \cite{Duke2000,Vilfan2003,Hilbert2013,Walcott2012,Erdmann2012,Erdmann2013,Erdmann2016}.
One important aspect is the
realization that myosin II acts as a catch bond, which means that bond lifetime is increased under mechanical force \cite{Veigel2003,Luo2012}.
This leads to accumulation of myosin to stressed parts of the actin network \cite{Luo2013}. Earlier we have incorporated the catch bond character of myosin II
in a master equation approach for minifilaments and showed that it can explain many aspects of cellular mechanosensitivity \cite{Albert2014}.

While the force generating aspect of myosin II has been studied and modeled in great detail, the literature describing the dynamic self-assembly of myosin II minifilaments is less developed. For myosin II minifilaments from the amoeba Dictyostelium,
a very detailed model has been developed, that however incorporates some biological details that
do not necessarily apply to other minifilament systems \cite{Mahajan1996,Ren2009,Luo2012,Luo2013,Luo2015}.
Here we aim at a more generic model in the spirit of the aggregation-fragmentation theory by
Smoluchowski \cite{Smoluchowski1917} and Becker and Döring \cite{Becker1935}, which is the standard model for assembly processes.
Our starting point is the observation that non-muscle myosin II minifilaments from human cells assemble to a stereotypic size of 28 to 30 molecules,
corresponding to a linear size around 300 nm \cite{Billington2013a,Niederman1975}.
Thus, they are an example for molecular assemblies of well-defined size, similar to e.g. virus capsids,
whose assembly has been modeled before in great detail \cite{Zlotnick1994,Rapaport2004,
Sweeney2008,Hagan2014}. The very regular architecture of the
virus capsids could be determined by electron microscopy, which motivated self-assembly models that
were built upon the neighborhood relations of the constituents of the capsid.
It has been argued that assembly of finite-sized complexes works best if the cluster size distribution is relatively flat
with peaks only for the monomer and complete complexes \cite{Boettcher2015}.
Other examples for self-assembling protein complexes that have been modeled include clathrin coats, adhesion complexes,
cytoskeletal fibers and chromatin \cite{Hafner2019}.

In order to model myosin II self-assembly in detail, we use the observation that
non-muscle myosin II assembles into bipolar filaments of approximately 30 proteins by electrostatic interactions of the coiled-coil tail domain,
where electric charges are periodically arranged and support both parallel and anti-parallel alignment of rods \cite{Ricketson2010}. Binding energies have been estimated to be about $35\,k_B T$ at zero ionic strength, however, due to the low screening length in cytoplasma ($\sim 1\,$nm at $100\,$mM NaCl) for physiological conditions, we expect and employ much lower binding energies in our model. Different options for rod arrangement within a bipolar filament have been proposed for muscle myosins of various species \cite{Squire1973, Chew1995}. The three-dimensional structures of the side regions of bipolar filaments of muscle cells from different species have been reconstructed from cryo-electron microscopy images with a resolution of $\sim 2\,$nm \cite{Woodhead2005, Kensler2012}. This quasi-atomic resolution has been able to be achieved using the known helicity of and periodicity within the side regions of the muscle bipolar filament. In these region myosin heads project out from the core, which is made up of the myosin tails, at equidistantly recurring axial levels, so called crowns \cite{Squire2009}. Between two subsequent crowns there is typically a well defined axial twist that varies between species. Potentially due to the missing spatial periodicity in the bare zone, until now it has only been possible to reconstruct the bare zone of bipolar filaments with a resolution of $\sim 5\,$nm which does not suffice to identify individual myosin tails \cite{AL-Khayat2010}. The authors could nevertheless show that the bare zone consists of multiple protofilaments interacting with each other. Here we will use this molecular information to develop an assembly model that takes this known molecular information into account, but on the other side is generic enough to describe myosin II minifilaments from different species. We then couple it to our crossbridge model for force generation and analyze the combined model in great detail.
Finally we will discuss its relation to experimental data.

This article is organized as follows. We first introduce our model as a graph. Growth of a minifilament is identified
with increasing occupancy of the nodes of this graph. For a given cluster, we then assume actin binding and force
generation through motor cycling. We analyze the dynamics of the combined model and identify steady states.
We find that at a certain monomer concentration the relaxation time increases dramatically. We explore the equilibrium properties of the model as a function of applied force and monomer concentration around this concentration, revealing that already unloaded actin facilitates minifilament assembly, with applied force enhancing this effect. In addition we produce fluorescence recovery after photobleaching (FRAP) trajectories which can be compared to experiments.

%% file: model.tex
\section{Models \& Methods} \label{sec:models}

\subsection{Minifilament Organization} \label{sec:detailedorganization}

\begin{figure*}[htbp]
\centering
\includegraphics[width=\textwidth]{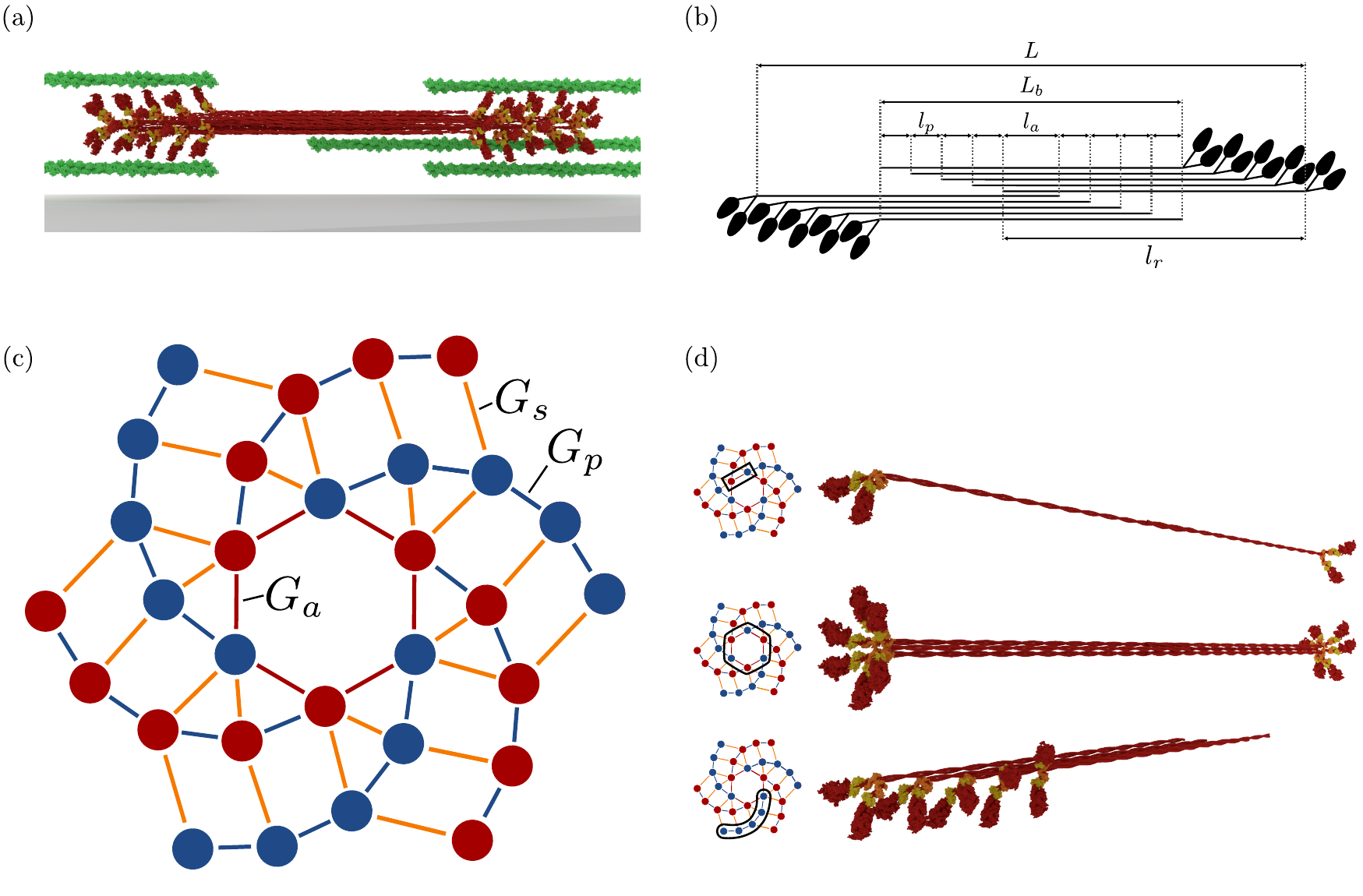}\\
\caption{Assembly model. (a) Artistic 3D rendering of a myosin minifilament that contracts actin fibres. (b) Schematic representation of a slice consisting of two anti-parallel protofilaments with indicated lengths (minifilament length $L$, bare zone length $L_b$, parallel stagger $l_p$, anti-parallel stagger $l_a$ and rod length $l_r$). (c) The graph on which the assembly occurs. The red lines in the middle represent strong interactions between anti-parallel myosin rods, the blue lines represent interactions caused by a favorable parallel overlap of myosin rods and the orange lines represent weak anti-parallel interactions. The red and the blue circles represent sites with opposing myosin heads. (d) Artistic representation of intermediates (right), with corresponding regions of the graph indicated (left). (top) Initial nucleation seed of two anti-parallel molecules. (middle) Inner core of the minifilament containing three of the nucleation seeds. (bottom) One protofilament.}
\label{fig:model}
\end{figure*}

Fig.~\ref{fig:model}a shows an artistic representation of a myosin II minifilament that is contracting opposing actin filaments.
The number of myosin II molecules in a minifilament has been estimated to be
between 28 and 30 \cite{Niederman1975,Billington2013a}, from which we take the later
value, because it allows for a more symmetric cluster architecture. 
Each myosin is a hexamer comprised of two myosin heavy chains, two essential light chains and two regulatory light chains.
The heavy chain globular region (i.e. the myosin head) can bind to actin filaments and displace them by undergoing
a powerstroke in the neck region behind the head region.
The two heavy chains form a long and relatively stiff rod due to hydrophobic-hydrophilic interactions.
To the outside, this rod carries a very specific pattern of charged amino acids which leads to
favorable interactions with other myosin rods at well-defined staggering distances \cite{Squire1973,Ricketson2010,Billington2013a}.
The most important one seems to be the anti-parallel overlap at $l_a = 45$ nm, which establishes the
basic bipolar structure of the minifilament. The most favorable parallel staggers are at $14.3$ nm and $43$ nm.
Here we focus on the first one, $l_p = 14.3$ nm. Fig.~\ref{fig:model}b shows a schematic two-dimensional representation
of the most likely arrangement of myosin II rods in a slice through the minifilament given these two prominent staggers.
With the rod length $l_r = 160$ nm, the overall minifilament length is $L = 2 l_r - l_a = 275$ nm and
the length of the bare zone (no myosin heads) is $L_b = 2 l_r - l_a - 8 l_p = 160$ nm, in good agreement with
electron microscopy data \cite{Billington2013a}. Note that three such slices have to be combined to give
the full minifilament with 30 molecules.

In order to represent the full three-dimensional structure of the minifilament, we represent it by the
graph shown in Fig.~\ref{fig:model}c. Here the two opposing directions of the rods are represented by two different colors for the nodes.
The core of the filament is defined by six rods forming a hexagon, with three rods from each direction.
They are held together by the anti-parallel overlap with staggering length $l_a$ and we assign
a binding energy of $G_a$ to this kind of bond.
From each of the six rods in the core, one string with four additional rods of the same orientation
spirals out to the periphery. These five rods define the five crowns and together our graph
contains the $6 \times 5 = 30$ molecules assumed in our model. Note that two neighboring spirals
together form the slice shown in Fig.~\ref{fig:model}b. The spiraling rods are held together
by the parallel stagger with $l_p$ and we assign a binding energy $G_p$ to these bonds. Because
intermediates with both anti-parallel and parallel staggers have been observed and in the
absence of further information, here we assume that $G_a$ and $G_p$ have similar values. Note that
the three-dimensional structure does not change the linear lengths $L$ and $L_b$ for the minifilament
and the bare zone given above.
Finally we note that our graph from Fig.~\ref{fig:model}c requires anti-parallel rods of not so favorable staggers to be in close proximity. Although not directly observed experimentally yet, these interactions must be present in order to fulfill the geometrical constraint that the bare zone is roughly six times as thick as the diameter of one myosin tail while maintaining an architecture which is organized from a core that is located in the center of the minifilament. We note that a central core and thereby very accessible side regions explain the relatively fast exchange times
that have been measured with FRAP \cite{Hu2017, Shutova2017} and also the dynamic rearrangements
of minifilaments observed in live cell microscopy with structured illumination \cite{beach_actin_2017}. We
assign a relatively low binding energy $G_s$ to this kind of bonds.

Although our model is a strong simplification, it captures all the geometrical
properties known from the literature. We note that
it is highly likely that real minifilaments are more disordered than assumed here.
For example, we do not expect all six rods in the core to be exactly aligned,
because they form a tight bundle in which also next-nearest neighbors
are relevant and which might use some of the other staggers known for myosin \cite{Ricketson2010}.
It is also known that different species form different staggers and have different
rod architectures. The graphical model suggested here should be considered to
be a consensus architecture that captures most of the known general features
of myosin II minifilaments.

\subsection{Minifilament Assembly} \label{sec:detailedassembly}

We now use the graph introduced in Fig.~\ref{fig:model}c to define the minifilament growth dynamics.
Starting from one myosin molecule in the core, the minifilament most likely polymerizes by
recruiting new myosin molecules onto neighboring sites. Thus the growth dynamics can
be represented by populating more and more of the nodes of the graph. Fig.~\ref{fig:model}d
shows different intermediates of the assembly process, both as subsets of the graph and
as artistic representations in space. We assume that association is
diffusion-limited, with a rate $k_\text{on}$ that does not depend on the binding energy
gained, but is proportional to the concentration of
myosin molecules. Dissociation corresponds to turning an occupied site into an unoccupied one.
Assuming detailed balance, this occurs with a rate
\begin{align}
k_\text{off}=k_{\text{off}}^{0} \exp \left(- \frac{n_s G_s + n_p G_p + n_a G_a}{k_B T} \right) \label{eq:koff}
\end{align}
that is dependent on the number of each particular bonds ($n_s$, $n_p$, $n_a$) that are broken due to the removal of the dissociating myosin II molecule.
If a myosin dissociates from the minifilament such that two separate patches are generated,
we remove the patch that does not contain the central region of the graph. Our growth model
is now complete and can be simulated using the Gillespie algorithm for reaction kinetics \cite{Gillespie1976}.

\subsection{Crossbridge Model}

\input{parameters}

\begin{figure*}
\includegraphics[width=\linewidth]{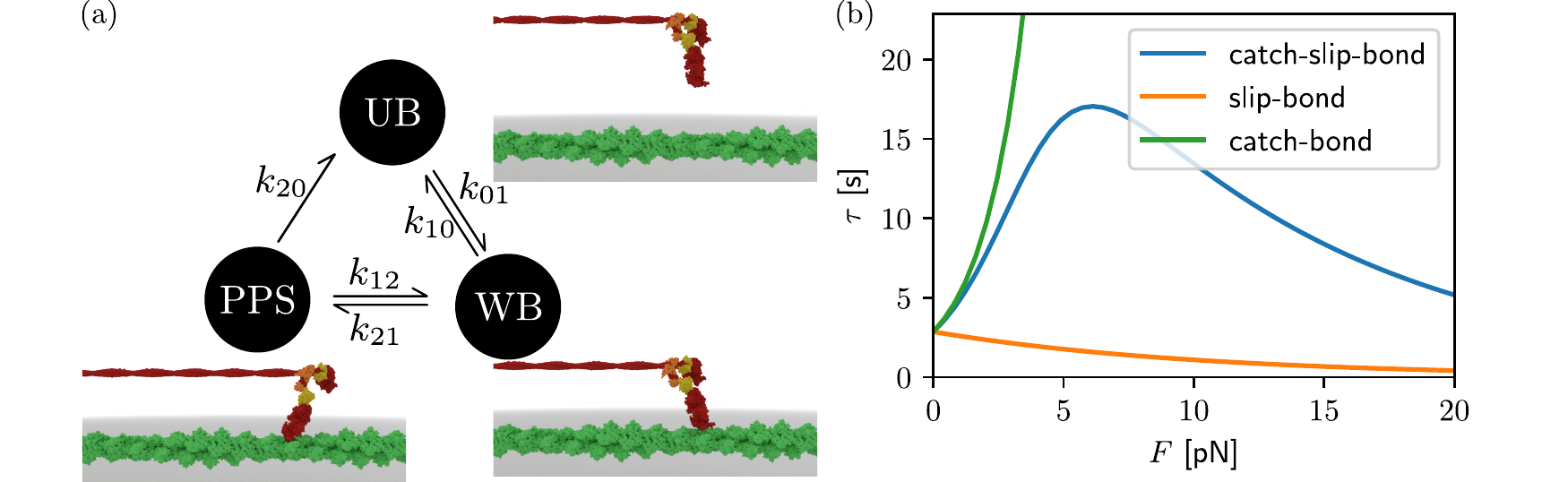}
\caption{Parallel Cluster Model (PCM) for force generation. (a) The PCM considers the three most important states of the
crossbridge cycle. The reaction rates $k_{01}$ and $k_{10}$ are constant, while the rate $k_{20}$ depends on force as given in \eqref{eq:catchslip}. The rates $k_{12}, k_{21}$ are high compared to the other rates. (b) Mean dwell time of a single myosin head on actin assuming catch-slip bond ($\Delta_c=0.92$), a pure slip bond ($\Delta_c=0$) and a pure catch-bond ($\Delta_c=1$).}
\label{fig:catchslipdwell}
\end{figure*}

The crossbridge cycle of a single myosin II protein is modeled according to the Parallel Cluster Model (PCM) \cite{Erdmann2012, Erdmann2013}.
In the PCM, the crossbridge cycle is described by a three-state system as depicted schematically in Fig.~\ref{fig:catchslipdwell}a.
The first state of the PCM is the unbound state (UB) of myosin. From there a myosin head can bind to actin into the weakly bound state (WB). Now the lever arm can swing backwards which reversibly transitions the myosin head to the post-powerstroke state (PPS). This transition is very fast (ms). Finally, from the PPS state myosin can unbind from actin via two different reaction paths, namely the catch-path and the slip path. The reaction rate along the catch-path decreases exponentially with increasing force, while along the slip-path it increases exponentially. The model summarizes these effects into a cumulative rate that depends on the force that the myosin-actin bond retains, i.e.
\begin{align}
\begin{split}
k_{20}(F) &= k_{20,\,0} \left[ \Delta_c \exp \left(-\frac{F}{F_c}\right)\right. \\
 &\quad + \left. (1- \Delta_c)  \exp \left( \frac{F}{F_s} \right) \right]\, ,
\end{split} \label{eq:catchslip}
\end{align}
where $\Delta_c$ is the fraction of myosin heads that use the catch-path to unbind at zero force, $F_c$ and $F_s$ are the critical forces for the catch-path and the slip-path, respectively, and $k_{20,\,0}$ is the rate at zero force. The inverse of the rate, i.e. the mean dwell time, is shown in Fig.~\ref{fig:catchslipdwell}b.\\
When part of an ensemble is retaining a force, it is assumed that, by consecutive unbinding and rebinding of the heads, the strain of all motors that are in the same mechanochemical state is the same. Thus, the strain only depends on the current state of the ensemble (i.e. how many motors are in each state of the crossbridge cycle) and not on the history of the filament. In this manner the model describes an ensemble of $N$ motors where $i \le N$ motors are in an actin bound state and $j \le i$ motors have performed the powerstroke. Hereby it is possible to calculate the strain $x_{ij}$ of the weakly bound motors when the cluster is balancing against an external force $F_\text{ext}$ yielding $x_{ij}=(F_\text{ext}-jkd)/ik$, where $k$ is the spring constant of the neck linkers and $d$ the length of the powerstroke.

The high transition rates between the PPS state and the WB state compared to the unbinding rates allow to maintain a local thermal equilibrium (LTE) between the two bound states. The probability for $j$ motors being in the PPS state, when $i$ are bound, follows the Boltzmann distribution $p(j|i)=\exp(-E_{ij}/k_B T)/Z$, with the partition sum $Z$. The energy $E_{ij}=E_\text{el} + j E_\text{pp} + E_\text{ext}$ is the sum of the elastic energy $E_\text{el}= k [(i-j)x_{ij}^2 + j(x_{ij}+d)^2]/2$ stored in the neck linkers, the free energy bias towards the PPS state $E_{pp} \approx -60\,\text{pN}\,\text{nm}$ and the contribution of any conservative external force field $E_\text{ext}$.  For a non-conservative constant force -- as discussed here -- $E_\text{ext}=0$. \\
LTE of the bound states allows us to average over all possible numbers of motors $j$ in the PPS, thus making it possible to describe the probability of $i$ motors bound to actin in the one-step master equation
\begin{align}
\begin{split}
\frac{\text{d}}{\text{d} t } p_i &= r(i+1) p_{i+1} + g(i-1) p_{i-1}\\
 &\quad - [r(i) + g(i)]p_i \, .
\end{split}
\end{align}
As $N-i$ motors can bind, the binding rate $g(i)$ is given by $g(i)=(N-i)k_{01}$. Unbinding is possible from the WB state and the PPS state such that the rate reads $r(i,j)=(i-j)k_{10} + jk_{20}(f(i,j))$, where $f(i,j) = (F_\mathrm{ext}-dk(i-j))/i$ is the force that is retained by one motor in the PPS state.
Averaging over $j$ yields $r(i)=\sum_j p(j|i)r(i,j)$. In the case of constant non-conservative forces, this sum has been found to be approximated well by $r(i)= r(i,i)$ \cite{Erdmann2013}. The model depends strongly on the chosen rates. Here we use the rates that we have previously used to study non-muscle myosin IIB \cite{Erdmann2016}, which is considered to be the main isoform responsible for maintaining long lasting forces \cite{Shutova2018}.

\subsection{Coupling of Self-Assembly and Force Generation}

Each occupied site of the self-assembly model can be in one of the three states of the crossbridge model.
The two sub-ensembles with the different orientations (blue and red in the graph) work against each other
in a tug-of-war situation which has been modeled before with the PCM for fixed minifilament sizes \cite{Albert2014}.
This implies that force can be generated only if both sides are attached to actin.
Here we assume that for each two-headed myosin molecule, only one head can be
active at a given time, as experiments have suggested that one of the two heads mainly
optimizes the force generating action of the other, while not being active itself \cite{Tyska1999, Kad2003}. 
Thus from the $60$ heads, only $30$ are considered in our model.
To complete the model, we now have to couple the minifilament to a specific mechanical environment.
Here we choose to work with a constant force ensemble, in contrast to an elastic environment with
own stiffness. In this way, we can avoid any dependence of our model on neck linker stiffness, whose effective
value is known to depend on context \cite{Veigel2003,Kaya2010}. Earlier, neck linker stiffness values have been used that lead
to strong occupancy of the PPS states  \cite{Erdmann2012,Erdmann2013,Albert2014}.
In the combined model, actin-bound motors cannot dissociate from the ensemble directly,
but must first unbind from actin by going into the UB state. This makes dissociation of the actin bound motors a two-step process
that depends on features of both models.

\subsection{Mean-Field Theory}\label{sec:deterministic}

In order to obtain an intuition for the behavior of the system it is instructive to coarse grain the assembly model to one variable.
We consider one side of the minifilament and denote its size with $N$. A monomer addition scheme for polymerization means
\begin{align}
H_N + H_1 \xrightleftharpoons[\alpha_{N+1}]{\beta} H_{N+1}\, ,
\end{align}
where $H_1, H_N, H_{N+1}$ represent monomers, $N$-mers and $(N+1)$-mers, respectively, and $\beta$ and $\alpha_{N+1}$ are the association and dissociation rates for the half-filament.
The equilibrium size distribution for this model is solved recursively via detailed balance:
\begin{align}
p_{N+1}=\frac{\beta}{\alpha_{N+1}}p_N \, .  \label{eq:alphaN}
\end{align}
If we now require the dissociation rates $\alpha_N$ of the monomer addition scheme to be such that the equilibrium size distribution of one side of the minifilament assembly model from section \ref{sec:detailedassembly} is reproduced, equation \eqref{eq:alphaN} provides a conditional equation for the dissociation rates $\alpha_N$ if $\beta$ is given. The rate $\beta$ however is not known since the total association rate depends on the current assembly state of the minifilament. We assume $\beta=3 k_\text{on}$, as each side of the graph is made up of three protofilaments. Additionally, matching with the assumption that actin-bound motors cannot dissociate from the ensemble, the dissociation rate has to be weighted by the fraction $(N-i)/N$ of motors in the UB state.

Now, the state of the filament can be projected to two integers per side of the filament, the cluster size $N$ and the number of actin bound motors $i$. The master equation for one side of the filament is
\begin{align}
\begin{split}
\frac{\text{d}}{\text{d}t} p_{N, i}= &- \left(\alpha_N\frac{N-i}{N} + \beta\right) p_{N,i} \\
\quad &+ \alpha_{N+1} \frac{N+1-i}{N+1} p_{N+1, i}\\
\quad &+ \beta p_{N-1, i}\\
\quad &- (r_i + g_{N, i})p_{N, i} \\
\quad &+ r_{i+1}p_{N, i+1} + g_{N, i-1} p_{N, i-1} \, .
\end{split}\label{eq:master}
\end{align}
From the master equation \eqref{eq:master} it is possible to construct a mean-field description. Starting from
\begin{align}
\begin{split}
\left \langle \frac{\text{d}}{\text{d} t} N \right \rangle &= \sum_{N, i} N \frac{\text{d}}{\text{d} t} p_{N,i}\\
\left \langle \frac{\text{d}}{\text{d} t} i \right \rangle &= \sum_{N, i} i \frac{\text{d}}{\text{d} t} p_{N,i}
\end{split}
\end{align}
and shifting summation indices and Taylor expanding around ($\langle N \rangle, \, \langle i \rangle$) 
yields
\begin{align}
\begin{split}
\langle \dot{N} \rangle &= \beta - \alpha(\langle N \rangle) \frac{\langle N \rangle - \langle i \rangle}{\langle N \rangle} \\ &\quad + \mathcal{O}(\sigma_N^2 + \sigma_i^2 +  \text{cov}(N,i))\\
\langle \dot{i} \rangle &= (\langle N \rangle - \langle i \rangle)k_{01} - \langle i \rangle k_{20}(F/ \langle i \rangle) \\ &\quad + \mathcal{O}(\sigma_N^2 + \sigma_i^2 + \text{cov}(N,i))\, .
\end{split} \label{eq:meanfield}
\end{align}
It is possible to compute the time development of the second central moments. These however depend on third central moments, resulting in a closure problem. In the following the second central moments are dropped for simplicity.

From equation \eqref{eq:meanfield} it is possible to calculate the two nullclines of the system
\begin{align}
\begin{split}
\langle N \rangle_{\langle \dot{i} \rangle =0}&=\langle i \rangle \left(1+\frac{k_{20}(F/\langle i \rangle) }{k_{10}}\right)\\
\langle i \rangle_{\langle \dot{N} \rangle =0}&=\langle N \rangle \left(1 - \frac{\beta}{\alpha(\langle N \rangle)} \right)\, .
\end{split} \label{eq:nullclines}
\end{align}

\subsection{FRAP-experiments}\label{sec:frapmodel}

The presented model allows for performing \textit{in silico} FRAP experiments by associating another Boolean variable to every occupied site that indicates whether the associated myosin is fluorescently labeled. By starting the Monte Carlo simulation from a non-fluorescent state drawn from the equilibrium distribution and filling up holes that form after dissociation of one molecule with new, fluorescent myosin proteins, FRAP traces can be obtained by calculating the time course of the ensemble average of the number of fluorescently labeled sites.

%% file: parameters.tex
\begin{table}[bt]
\caption{Parameters used in the simulation of minifilament assembly and force generation.}
\label{tab:param}
\begin{ruledtabular}
\begin{tabular}{l|ccc}
Parameter & Symbol & Value & References\\\hline
Transition &$k_{01}$ & $0.2$ & \cite{Stam2015} \\ 
rates $[ $s$^{-1}]$    &$k_{20,\,0}$ & $0.35$ & \cite{Stam2015} \\  
                 &$\Delta_c$ & $0.92$ & \cite{Stam2015} \\ 
Force scales     &$F_c$ & $1.66$ & \cite{Stam2015} \\ 
$[$pN$]$         &$F_s$ & $10.35$ & \cite{Stam2015} \\ 
Energy scales    &$G_{a}$ & $3$ & our estimate \\ 
$[k_B T]$        &$G_{p}$ & $3$ &  our estimate \\ 
                 &$G_{s}$ & $1$ &  our estimate \\
\end{tabular}
\end{ruledtabular} 
\end{table}

%% file: results.tex
\section{Results} \label{sec:results}

\begin{figure*}[htb]
\includegraphics[scale=1]{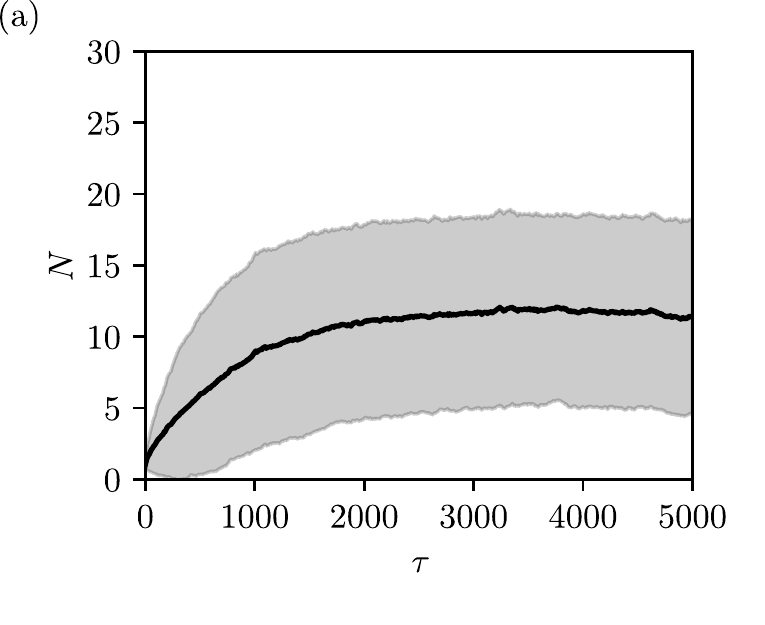}
\includegraphics[scale=1]{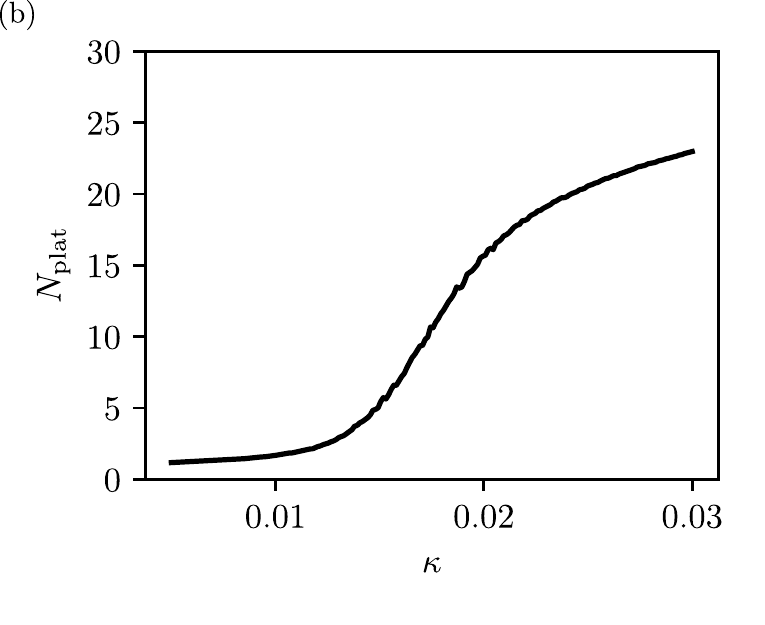}\\
\includegraphics[scale=1]{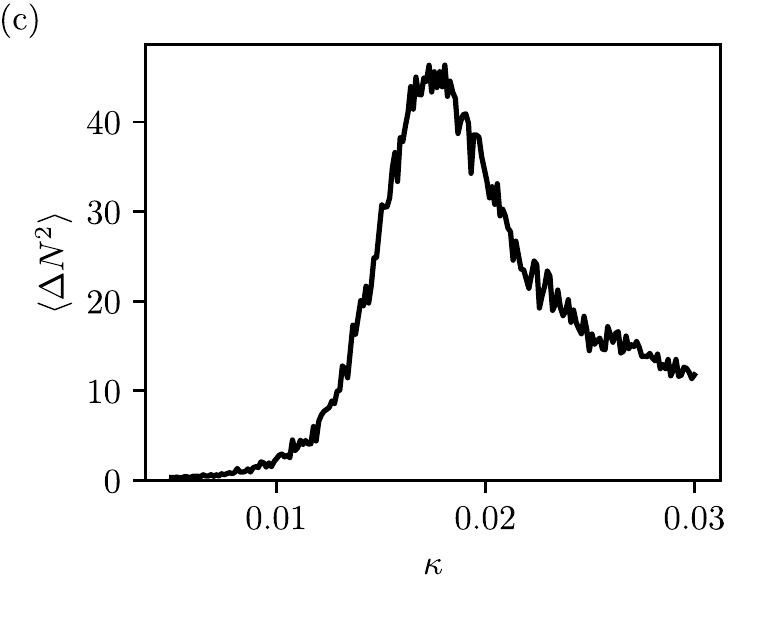}
\includegraphics[scale=1]{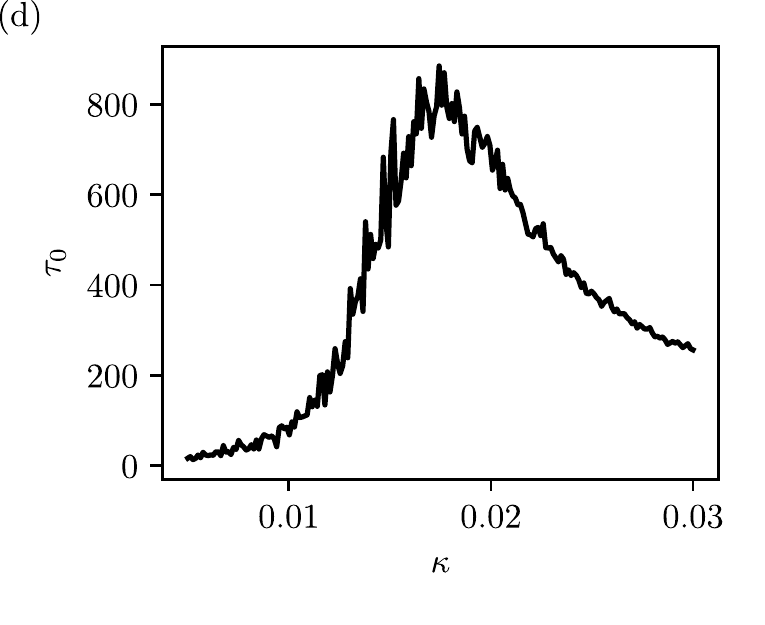}
\caption{Assembly dynamics. (a) Time course of the mean minifilament size for dimensionless
association rate $\kappa = 0.018$ (black line) with standard deviation (grey area).
(b) Mean number of assembled myosins as a function of $\kappa$. (c) Variance of $N$ as a function of $\kappa$. A peak at $\kappa_c \approx 0.018$ indicates the transition between partially and fully assembled minifilaments. (d) Relaxation time of the minifilament
as a function of $\kappa$ as obtained from a saturating exponential fit to the mean size of an assembling cluster.}
\label{fig:timecrit}
\end{figure*}

\subsection{Assembly dynamics}\label{sec:assemblyrate}

We first discuss the assembly model based on the graphical model from Fig.~\ref{fig:model}c, that is we
do not consider yet the coupling to the motor model.
We simulated the mean number of assembled myosins $N$ (maximal value 30) for the model described in section \ref{sec:detailedassembly} using the Gillespie algorithm and the parameter values from Table \ref{tab:param}.
Fig.~\ref{fig:timecrit}a shows the mean trajectory and its standard deviation. We see that the
mean assembly dynamics can be described well by an exponentially saturating function $N_a \left(1-\exp (-\tau / \tau_0)\right)+1$, where $\tau = t k_\mathrm{off}^0$ is the dimensionless time. Note that the minimal cluster size at $\tau = 0$
has to be $1$.

Fig.~\ref{fig:timecrit}b shows the plateau value $N_\mathrm{plat}=N_a+1$ as a function of the dimensionless association rate
$\kappa=k_\text{on}/k_\text{off}^0$. One sees that
the larger the association rate, the more the mean size $N_\mathrm{plat}$ approaches the maximal value $30$,
and that the function has a hyperbolic character, indicating a crossover at the inflection point.
Fig.~\ref{fig:timecrit}c shows the variance of $N$, which has a clear peak
at a critical value $\kappa_c = 0.018$, indicating
a transition between partially and fully assembled minifilaments. Fig.~\ref{fig:timecrit}d shows the relaxation time $\tau$
as a function of association rate $\kappa$, which again has a clear peak at $\kappa_c = 0.018$ (with value $\tau \approx 800$).
We interpret these results as critical slowing down. Because the association rate $\kappa$ is proportional
to the myosin II concentration in solution, the critical association rate $\kappa_c$ corresponds
to a critical aggregation concentration (CAC). In the following, we will investigate our model around this critical point.

\begin{figure*}[htbp]
\includegraphics[scale = 1]{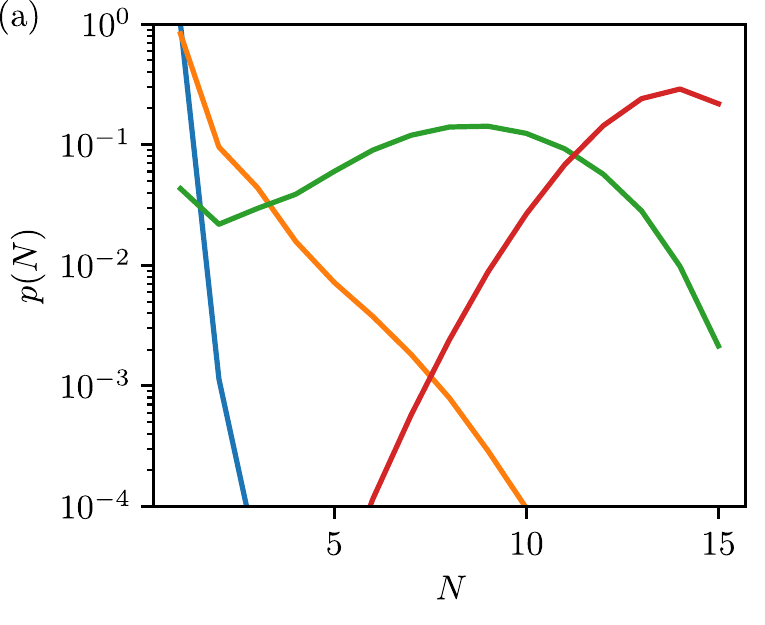}
\includegraphics[scale=1]{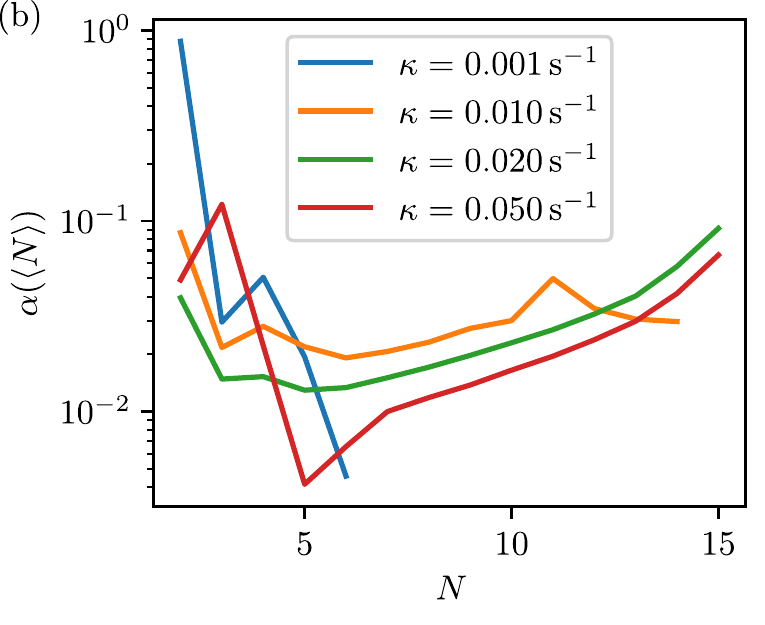}
\caption{Equilibrium distribution $p$ and dissociation rates $\alpha$ as a function of association rate $\kappa$.
(a) Equilibrium distribution $p(N)$ of the assembly model for different values of $\kappa$. (b) Resulting dissociation rates $\alpha(\langle N \rangle)$ (compare eqn. \eqref{eq:alphaN}) that are used in the mean-field model or in a coarse grained model.}
\label{fig:eqdist}
\end{figure*}

From the stochastic simulations, we can also obtain the full cluster size probability distribution.
From here on, we will still simulate the full minifilament,
but only show results for one half, because the two halves are statistically equivalent.
Thus from here on the maximal cluster size is $15$.
The size distribution for a half-filament is shown in Fig.~\ref{fig:eqdist}a. At association rates
below the critical value $\kappa_c$, the distributions are approximately exponential.
At the critical association rate, the distribution becomes very broad. Above the critical
value, a clear maximum emerges close to full assembly. As explained in section \ref{sec:deterministic}, from these
distributions one can calculate effective equilibrium constants (eq. \eqref{eq:alphaN}) that map the graph
model to a monomer addition scheme. Fig.~\ref{fig:eqdist}b shows the effective off-rate $\alpha_N$ obtained from
equation \eqref{eq:alphaN}. These are used in the mean field approach in the following.

\subsection{Steady state results}

\begin{figure*}[t]
\includegraphics{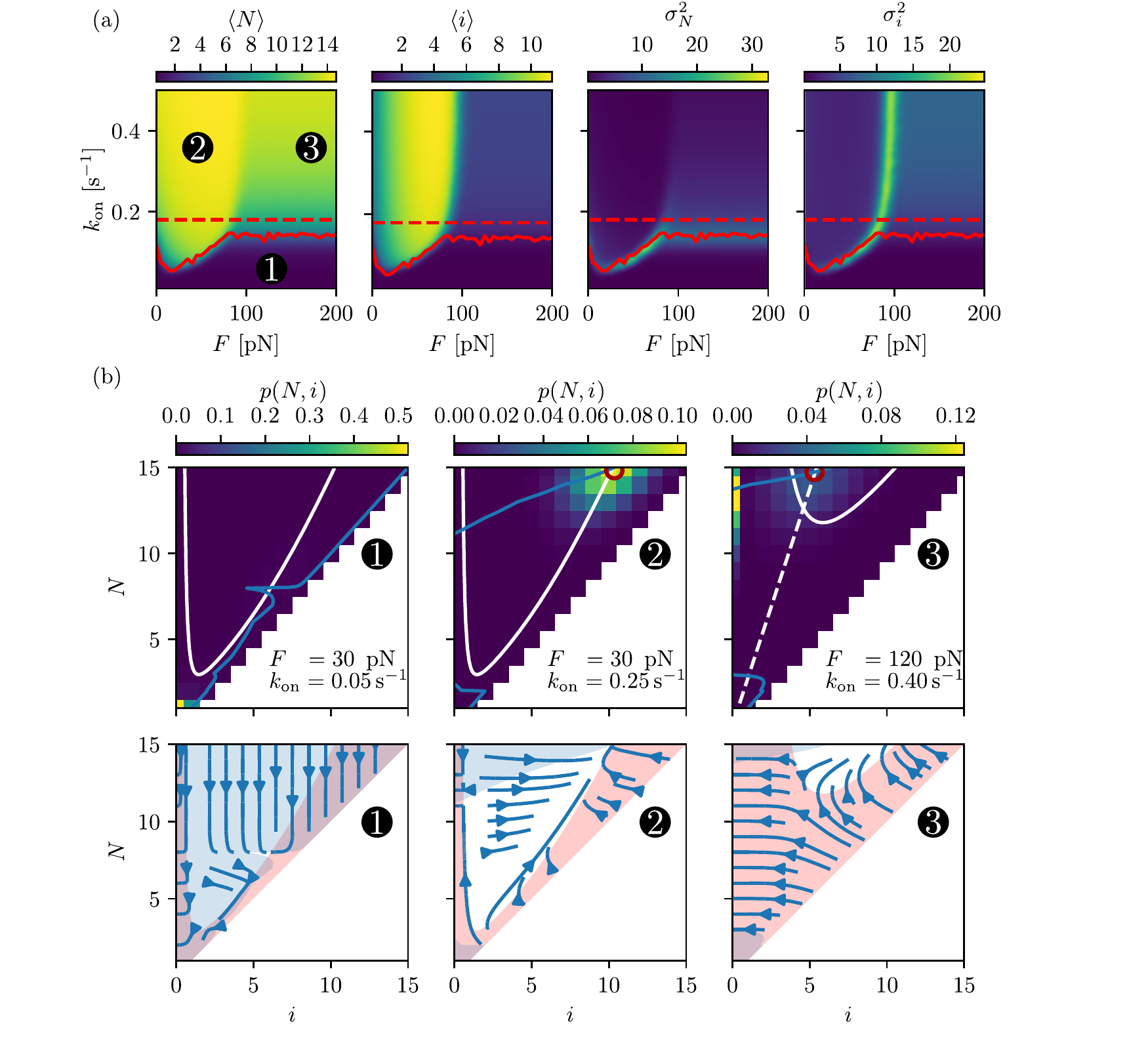}
\caption{Steady states. (a) Mean-values and variances of the full model for different forces $F$ and on-rate $k_\mathrm{on}$ ($k_\mathrm{off}^0=10\,\mathrm{s}^{-1}$). The red dashed line represents the critical on-rate $k_\text{on}$ for a cluster without actin, whereas the solid line represents the critical on-rate as a function of force. (b) Equilibrium distribution $p(i,N)$ (top row) and phase portrait (bottom row) in the 3 different regions. In the top row the solid white line depicts the nullcline of $i$ whereas the blue line depicts the nullcline of $N$. The dashed white line depicts the nullcline of $i$ at zero force. The red circles denote stable fixed points of the meanfield theory for the indicated force (middle) and zero force (right). The phase portraits (bottom row) illustrate how a change in force and on-rate affects the flow lines (blue) and the nullclines (red region $\langle \dot{i} \rangle < 0$, blue region $\langle \dot{N} \rangle < 0$).}
\label{fig:distphaseportrait}
\end{figure*}

We now investigate the full model that couples assembly and force generation.
Starting from here we stop using dimensionless quantities, since the dynamics of the myosin crossbridge cycle are experimentally measured for specific isoforms and we choose to study the effects on non-muscle myosin IIB, where the fraction of time a single myosin head is attached to actin (the so-called duty ratio) is comparatively high \cite{Erdmann2016}. In addition, we utilize the assembly rates documented in table \ref{tab:param} (justified later in section \ref{sec:frap}).
In order to obtain a complete understanding of our combined model, we investigate how the mean values of the number of assembled motors $N$ and the mean values of the number of bound motors $i$ of one side of the filament change with association rate $k_\text{on}$ and force $F$. In addition, we record the variances of these quantities, because this indicates transitions between
different regimes. The corresponding results are shown in
Fig.~\ref{fig:distphaseportrait}a. Here we also show the values of the critical association rate:
the solid and dashed lines show these transitions with and without motor cycle dynamics, respectively.
While the dashed line corresponds to the results from section \ref{sec:assemblyrate},
for the full model we numerically searched for the maximal relaxation time.
From Fig.~\ref{fig:distphaseportrait}a we see that there exist three different regimes (marked by
labels $1$, $2$ and $3$) which are separated
by small parameter regions with high variance in either the cluster size $N$ or the number of actin bound motors $i$.
The solid line for the critical values for the relaxation times nicely corresponds to the
transition region defined by the variance in $N$. The dashed line from the assembly model
is always higher, suggesting that actin binding lowers the CAC.

We now discuss the three different regimes identified in Fig.~\ref{fig:distphaseportrait}a in more detail.
The regime $1$ at low association rates $k_\text{on}$ is characterized by a small mean cluster size $\langle N \rangle$.
Due to low association, there are not enough monomers to support an assembled minifilament.
At higher association rate and up to medium forces, regime $2$ emerges, in which minifilaments are typically
assembled and attached to actin with both sides.
The border of this region to the prior one is convex, indicating that the catch-slip bond mechanism facilitates assembly under medium forces by increasing the amount of actin-bound myosin that is unable to dissociate from the minifilament. At higher forces and high on-rates, there is regime $3$, in which the minifilaments are typically assembled, however, the number of actin bound motors $i$ of the half-filament is reduced to half the value which one would obtain with $F=0\,\text{pN}$.
The underlying reason is that now the slip pathway dominates and therefore one half
of the minifilament unbinds, while the other side binds without force.

In order to understand why these three regimes form, in Fig.~\ref{fig:distphaseportrait}b we show
the probability distributions $p(N,i)$ for cluster size $N$ and bound motors $i$. In addition we show
the phase portraits of the deterministic (mean-field) system described in section \ref{sec:deterministic}.
As shown at the left side of Fig.~\ref{fig:distphaseportrait}b, the half-filament is of size $1$ at low monomer concentration and force. In the corresponding phase portrait one can see that the regime $1$ forms because the net flux of the system is always directed either towards lower size $N$ or lower actin-bound myosin heads $i$ which enhances each other in the model. Although the phase portrait shows a node at $(N\approx 8, i\approx 6)$, the proximity to a saddle makes it unstable to noise.

The regime $2$, in which both sides of the filament are attached, is shown in the middle of Fig.~\ref{fig:distphaseportrait}b and
is characterized by a stable fixpoint at large $N$ and $i$ with a large basin of attraction.
This leads to a maximum in the equilibrium distribution at the boundary $N=15$. In regime $2$, the
force is sufficiently small so that the motors are stabilized by their catch behavior.

In regime $3$, there are two populations: one at $i=0$ and the other distributed around the nullcline of $i$ at zero force. This indicates that the minifilament is not fully attached to actin, but only attached with one side and hence is not sustaining a force.
At this high level of force, the slip pathway dominates and
the mean field description fails, because it only describes one half-filament and assumes that the force can be applied.
This would not happen for a pure catch bond and our results for this case are shown as Fig.~1 of the supplement.
Then minifilaments can assemble at very low on-rate, just as long as the force is high enough. Additionally, there is no region where only one side of the filament is attached, but both sides are typically attached at the same time.

In summary, the force-dependence of the distributions indicates that with increased force, the probability for the system to be near the assembled maximum of the distribution increases. At high forces this probability decreases again. This behavior is illustrated in Fig.~2 of the supplement where the probability that the system is of size $N\geq 8$ is shown.

\subsection{Comparison with Experiment}\label{sec:frap}

\begin{figure*}[tb]
\includegraphics[scale=1]{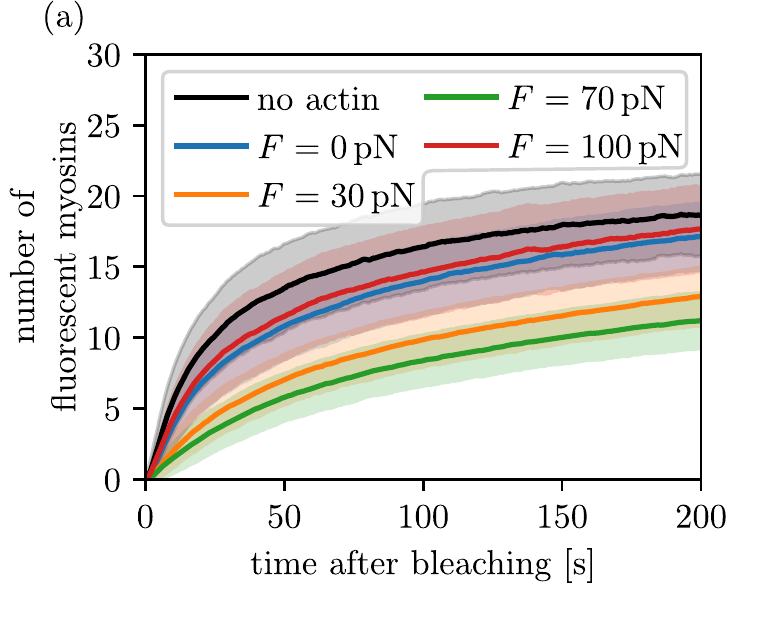}
\includegraphics[scale=1]{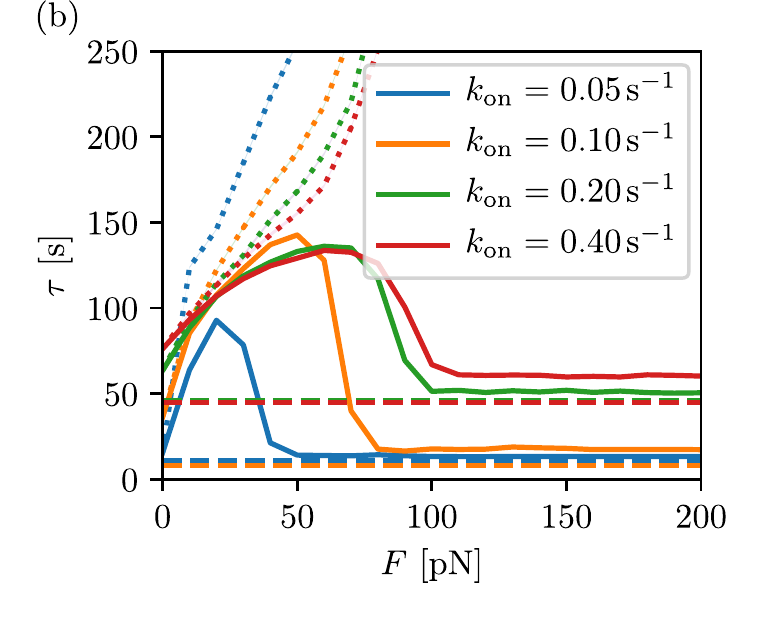}
\caption{(a) Time-dependent mean number of fluorescently labeled myosin proteins per minifilament starting from a non-fluorescent minifilament drawn from the equilibrium distribution at $k_\text{on} = 0.4\,$s$^{-1}$ for different forces and with the myosins heads blocked in the unbound state. The transparently colored regions denote the region of one standard deviation. (b) Results for fitting functions of type $N (1-\exp(- t/\tau) )$ to the mean number of fluorescently labeled myosins for different on-rates and forces. The dashed lines indicate the fluorescence recovery if the motor cycle is turned off, which is always significantly faster, even at forces where the minifilament is not bound to actin on both sides but typically only on one. The dotted lines indicate the result when using pure catch-bonds.}
\label{fig:fitresults}
\end{figure*}

As described in section \ref{sec:frapmodel}, by using the proposed model, it is possible to predict trajectories of FRAP experiments from the model. We investigated the effect of different forces with or without the crossbridge cycle, of which the latter mimics myosins heads that are blocked in the unbound state (experimentally this can be achieved by using the
pharmacological inhibitor blebbistatin \cite{Kovacs2004}). Fig.~\ref{fig:fitresults}a shows the mean number of fluorescent proteins in a minifilament $ \langle N(t) \rangle$ starting the dynamical self-assembly simulation with a non-fluorescent minifilament drawn from the appropriate equilibrium distribution. Similar to the fluorescence intensity in FRAP experiments, $\langle N (t)\rangle$ is a saturating and monotonously increasing function of time that can be described by a saturating exponential.

When fitting an exponential function of type $N_a (1-\exp (-t/\tau) ) $ to the fluorescence recovery traces at different forces and different on-rates close to the critical on-rate for the minifilament without actin, we choose $k_\text{off}^0 = 10\,$s$^{-1}$ such that we obtain values close to the recovery times measured in cells \cite{Hu2017, Shutova2017} (see Fig.~\ref{fig:fitresults}b). We note that the recovery times calculated here are on the lower end of the wide spectrum of reported experimental values, indicating that $k_\text{off}^0$ should be seen as an upper bound. For increasing force the fluorescence recovery time increases until it reaches a maximum at around $80\,\text{pN}$ from where it drops down to a constant value. This constant value is always higher than the fluorescence recovery time for minifilaments without actin, underlining once again that at very high forces one side of the minifilament is attached. If pure catch bonds are used to simulate the motor dynamics, the fluorescence recovery time rises monotonically with force, underlining that the drop we observe at intermediate forces for the catch-slip bond occurs due to the instability of slip bonds beyond a certain force. With our choice for the value of $k_\text{off}^0$, we revisit Fig.~\ref{fig:timecrit}d, which indicates the maximum relaxation time is $t_0 \gtrsim 80s$, consistent with light scattering measurements in \textit{in vitro} assembly assays \cite{Billington2013a} ($t_\text{exp} \approx 580\,$s).

%% file: outlook.tex
\section{Discussion}

In this paper, we have proposed an assembly model of myosin II filaments, that explains the mechanosensivity of myosin II self-assembly by coupling assembly and motor activity in one model. In particular, we suggested a graph representing
the consensus architecture of human myosin II minifilaments. Although myosin II minifilaments
tend to differ in the details of their architecture from species to species, our approach is very
generic and does not depend much on the details of this graph (Figs.~3 and 4 in the supplement).
We investigate the dynamical model on this graph in a range around a critical aggregation concentration (CAC), which
we identified by critical slowing down. We identified and characterized three regimes. Finally
we performed fluorescence recovery after photobleaching (FRAP) simulations that yielded recovery times which we used to find plausible assembly rates.

It is a common feature of self-assembling systems that, as soon as the equilibrium concentration of free monomers is beyond a threshold, i.e. the CAC, it does not increase much anymore with added total monomer, since the added monomer goes mainly towards forming additional assembled structures \cite{IsraelachviliBook}. Myosin minifilaments do not form an exception here, as has been experimentally shown \cite{Liu2017}. This means, that a system of assembling myosin tunes itself, such that forming new filaments becomes very slow after the equilibrium concentration has been reached.
However, this is only valid for solutions without actin. Our simulation results suggest that
in contact with actin the assembly could be facilitated by coupling assembly to force generation. This leads to the CAC of minifilaments being lowered locally near actin filaments, in agreement with experimental observations \cite{Pardee1992}. If the solution can support the assembly of minifilaments already without actin, i.e. the concentration of monomers is near the CAC, minifilaments operating on actin might associate new myosin molecules with the critical association rate of the solution, which is well above the critical association rate of the minifilaments that are attached to actin. This mechanism in conjunction with regulation of the equilibrium between assembly-competent and incompetent myosin II \cite{Vicente-Manzanares2009} yields a system that can show a very dynamic response to change of external conditions. In addition, it explains the known mechanoaccumulative behavior of myosin II \cite{Schiffhauer2016}.

Since blebbistatin, an often utilized small molecule inhibitor of myosin, blocks the myosin II head domain in an actin-detached state \cite{Kovacs2004}, the assembly-enhancing effect of actin, that our model predicts, could be experimentally investigated using already available methods \cite{Hu2017}. Hence, the model assumption that a myosin II protein is not able to dissociate from its respective minifilament when its head is bound to actin can in principle be verified.

We were not able to fully explain the wide spectrum of recovery rates reported by changes in retained force alone. Instead, also other mechanisms will be involved. However the rates we extract are consistent with light scattering data from \textit{in vitro} assembly assays, where assembly has turned out to be slower by a factor of 7 than our lower bound. This seemingly large deviation should however be put into perspective by noting that \textit{in vivo} FRAP experiments \cite{Hu2017,Shutova2017} were used to obtain absolute rates, which then were used to compare the model to \textit{in vitro} experiments \cite{Billington2013a}, that in addition have been conducted at a $17\,$\textdegree C lower temperature.

Our model suggests that the strong force-dependence of minifilament self-assembly arises due to the catch-bond characteristic of unbinding myosin from actin after performing the powerstroke. This constellation, where self-assembly of a motor complex is markedly affected by the binding dynamics to its track, is not unique to myosin II. Another interesting
example is the bacterial flagellar motor (BFM), which in contrast to myosin II is a rotary motor, but similar to minifilaments is a complex with multiple load bearing elements (i.e.\ stators). It has been shown that
increasing load (i.e.\ torque) increases the amount of stators in the BFM \cite{Tipping2013}. Later studies have suggested this to be due to the dissociation rate of the stators decreasing with increased torque \cite{Berry2017,wadhwa_torque-dependent_2019}, i.e.\ the BFM also implements a catch bond which modulates self-assembly. The catch bond feature is also central to the function of actomyosin, where it modulates the transient response to mechanical stress and guides accumulation of myosin to stressed parts of the actin network \cite{Veigel2003, Luo2012}. We conclude
that the interplay of assembly and force generation described here might be at play in other protein clusters that have to function under mechanical load.

%% file: supplement.tex
\clearpage
\widetext
\begin{center}
\textbf{\large Supplemental Figures}
\end{center}
\setcounter{equation}{0}
\setcounter{figure}{0}
\setcounter{table}{0}
\setcounter{page}{1}
\makeatletter
\renewcommand{\theequation}{S\arabic{equation}}
\renewcommand{\thefigure}{S\arabic{figure}}
\renewcommand{\bibnumfmt}[1]{[S#1]}
\renewcommand{\citenumfont}[1]{S#1}

\begin{figure*}[htbp]
\centering
\includegraphics[width = 0.9\textwidth]{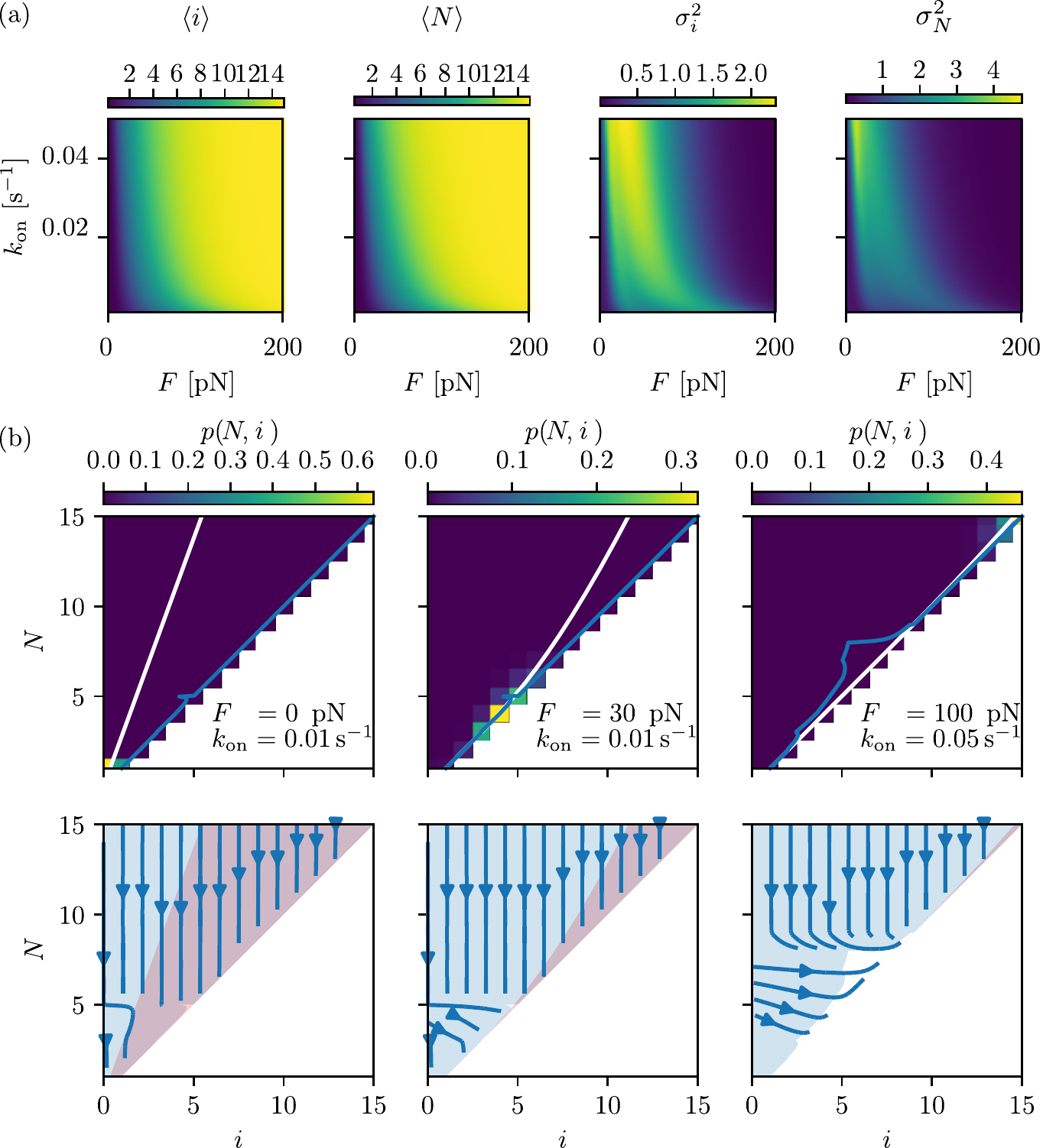}
\caption{Steady states. (a) Mean-values and variances of the full model for different forces and on-rates ($k_\text{off}^0=10\,$s$^{-1}$). Here the results are shown for catch bonds only. (b) Phase portraits for selected parameters for pure catch bonds. In this case force can be sufficient to assemble the minifilament even if the on-rates are very small.}
\end{figure*}

\begin{figure*}[t]
\centering
\includegraphics[scale = 1]{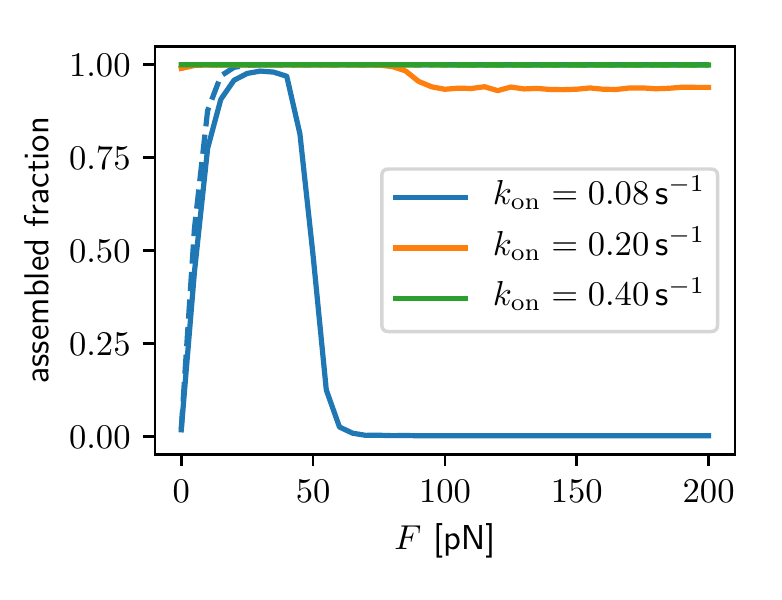}
\caption{The fraction of half-minifilaments with a size  $N \geq 8$ for different on-rates. Solid lines denote the simulations with catch-slip bonds, while dashed lines denote the ones with pure catch bonds (the orange and green dashed lines are both at 1).}
\label{fig:assembledfraction}
\end{figure*}

\begin{figure*}
\centering
\includegraphics[scale = 1]{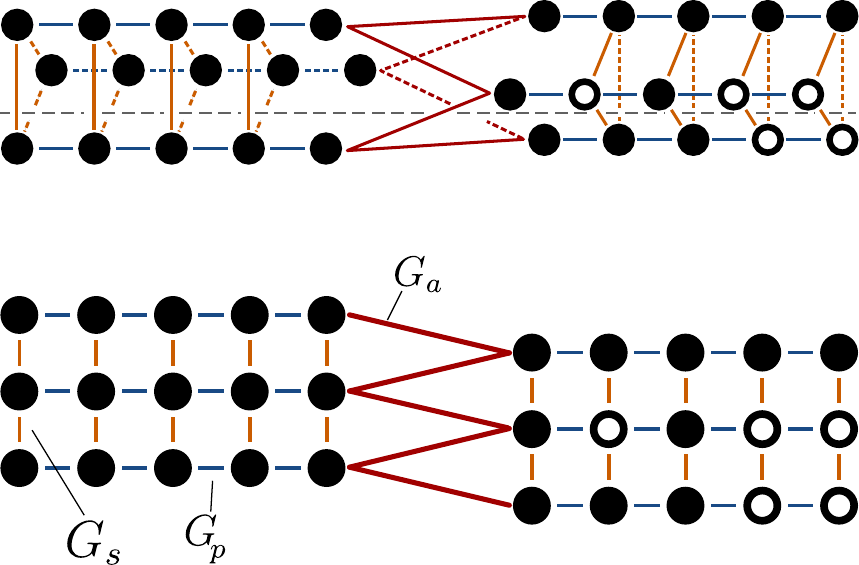}
\caption{Alternative graph topology, which yields the results shown in Fig. \ref{fig:altgraphres}. The top schematic illustrates the connectivity to be similar to the vertices of 2 prisms that have been axially rotated against each other by $180^\circ$. Cutting the bonds along the gray dotted line yields the unwrapped graph on the bottom.}
\label{fig:altgraph}
\end{figure*}

\begin{figure*}
\centering
\includegraphics[width = \textwidth]{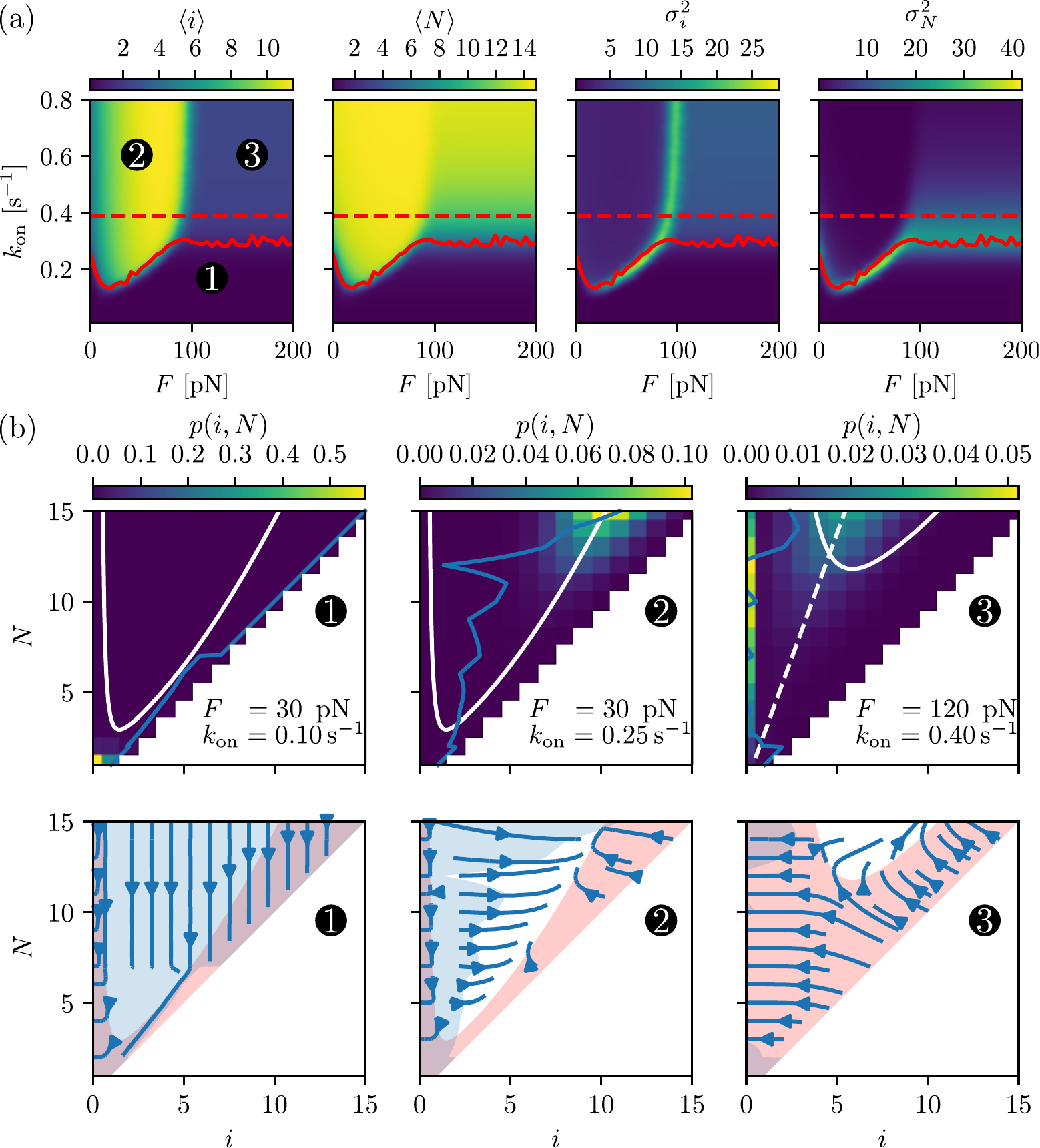}
\caption{Same quantities as Fig. 5 (main text) for the graph shown in Fig. \ref{fig:altgraph} ($k_\text{off}^0=20\,$s$^{-1}$),
giving similar results as the main model.}
\label{fig:altgraphres}
\end{figure*}